\documentclass[doublecol]{epl2}
\usepackage{graphicx}
\usepackage{amsmath}

\begin{document}

\title{The size, shape, and dynamics of cellular blebs}

\author{F. Y. Lim \inst{1} \and K.-H. Chiam \inst{1,2} \and  L. Mahadevan \inst{3}}
\institute{\inst{1} A*STAR Institute of High Performance Computing, 1 Fusionopolis Way, \#16-16, Singapore 138632, Singapore \\
\inst{2} Mechanobiology Institute, National University of Singapore, 5A Engineering Drive 1, Singapore 117411, Singapore \\
\inst{3} School of Engineering and Applied Sciences, Department of Physics, Harvard University, 29 Oxford Street, Cambridge, Massachusetts 02138, USA}
\shortauthor{F. Y. Lim \etal}


\abstract{
A cellular bleb grows when a portion of the cell membrane detaches from the underlying cortex under the influence of a cytoplasmic pressure.  We develop a quantitative model for the growth and dynamics of these objects in a simple two-dimensional setting. In particular, we first find the minimum cytoplasmic pressure and minimum unsupported membrane length for a stationary bleb to nucleate and grow as a function of the membrane-cortex adhesion.  We next show how a bleb may travel around the periphery of the cell when the cytoplasmic pressure varies in space and time in a prescribed way and find that the traveling speed is governed by the speed of the pressure change induced by local cortical contraction while the shape of the traveling bleb is governed by the speed of cortical healing.  Finally, we relax the assumption that the pressure change is prescribed and couple it hydrodynamically to the cortical contraction and membrane deformation.  By quantifying the phase space of bleb formation and dynamics, our framework serves to delineate the range and scope of bleb-associated cell motility and synthesizes a variety of experimental observations.}




\maketitle


\emph{Introduction ---}  Blebs are protrusions of a cell membrane driven by local variations in intracellular pressure induced by contractility and are commonly seen in many types of cells. Blebbing is a biological analogue of an elementary mechanical process --- the separation of a thin membrane from a bulk material. It is thus similar in many respects to the formation of a blister in a thin film adherent to a substrate, but also rather different in that it involves a number of active processes associated with active (and regulated) contractile stresses that drive the process, as well as active mechanisms associated with healing of these objects via forces at the boundary. Blebs have mostly been observed and studied during the processes of  apoptosis and cytokinesis, but recent studies have revealed cellular blebbing to be an important mechanism contributing to cell motility~\cite{Blaser-DevCell-2006,Charras-Nature-2005,Charras-molcellbio-2008,Langridge-ExpCellRes-2006,Yoshida-JCellSci-2006}.  For example,  migration of chemotactic germ cells in some systems is not powered by actin polymerization but by cytoplasmic streaming instead~\cite{Blaser-DevCell-2006}, while some tumor cells form bleb-like protrusions to migrate through three-dimensional matrices~\cite{Wolf-JCellBiol-2003}.   It has been further suggested that bleb protrusions can cooperate with lamellipodium-based protrusions to power cell motility~\cite{Charras-Nature-2005, Yoshida-JCellSci-2006}, and that some types of cells can switch between the two modes depending on their environment~\cite{Langridge-ExpCellRes-2006, Sahai-NatCellBiol-2003, Wolf-JCellBiol-2003, Friedl-NatRevCancer-2003}. Thus the focus on  actin polymerization-driven protrusions in lamellipodia and filopodia on cell motility must be complemented by considerations of  contractility/pressure induced  blebbing  before one can decipher the relative contributions of these modes for whole cell motility.  Understanding blebbing quantitatively is a first step in this process.

Here, we provide a theoretical model for the nucleation, expansion, retraction, and large scale movements of blebs.  A bleb is nucleated when the local hydrostatic pressure generated by cortical contraction  detaches the cell membrane from the cortex. Cytoplasmic pressure and flow then drives the detached membrane to expand, forming an expanding bleb that is not supported by the actin cortex except along its edge.  This expansion  is accommodated by further delamination of the cell membrane from the cortex, flow of lipid into the bleb through the bleb neck, and unwrinkling of excess folded membrane.  When the bleb expansion slows down, the actin cortex starts to reform underneath the bleb membrane.  In non-motile cells, myosin-driven contraction of the reformed cortex retracts the bleb, while in motile cells, contraction of the rear of a cell leads to a net movement of the cell body towards the leading edge where the bleb is formed. In addition, sometimes an asymmetric reformation of the actin cortex can lead to the circus movement of a bleb traveling around the periphery of a cell~\cite{Charras-BioPhysJ-2008}. During this last process, the actin cortex is reformed at one side of the bleb and its contraction pushes the cytosol to the other side of the bleb, which sometimes leads to further delamination of the membrane at the unconsolidated edge.

\emph{Model ---}
To quantify these phenomena, we focus on a simple 2 dimensional model of the dynamics of a cell, consisting of an active fluid --- a contractile cortex bathed in a Newtonian fluid, and surrounded by a membrane that is detachable from the actin cortex.   We assume that the membrane is one-dimensional and neglect its natural  curvature, which is reasonable for blebs that are small compared with the size of the cell.   In addition, we focus on a single bleb and thus neglect  bleb-bleb interactions, as well as the adhesion of the bleb to any external scaffolds or substrates. For a single bleb, denoting  the distance between the detached membrane and the cortex as $y(x)$ at the position $x$, the energy per unit width $E(x,y,y_x,y_{xx})$ of a piece of blistered membrane $y(x)$ is given by 
\begin{equation}
E = \int_{s_l}^{s_r} \left( \frac{B}{2}y_{xx}^2 + \frac{T}{2} y_x^2 + \frac{\kappa}{2} y^2 - p y - E_a \right) dx.
\end{equation}
where $(.)_x = d(.)/dx$, and the limits of integration $s_{l,r}$ define the left and right boundaries of the bleb, which are possibly dynamic.  The various terms in the integrand denote, respectively, the bending energy of the membrane with flexural rigidity $B$, the work done by the in-plane membrane tension  $T$, the binding energy of the adhesive bonds between the membrane and the cortex with spring constant $\kappa$, the work done by the cytoplasmic pressure $p$, and finally the adhesion energy between the membrane and the cortex. For consistency, the spring constant $\kappa$ must be related to the adhesion energy $E_a$ by $\kappa = 2E_a / l_c^2$, with $l_c$  the maximum length of the membrane-cortex adhesion bond, with the condition that when $y(x) \geq l_c$,  the adhesion bond at $x$ breaks.  The corresponding Euler-Lagrange equation for the shape of the bleb is then given by
\begin{equation}
B y_{xxxx} - T y_{xx} + \kappa y H[1-y/l_c] - p = 0,
\end{equation}
where $H[a]=1, a<0; H[a] = 0, a<0$ is related to the Heaviside function.  
In general, the cytoplasmic pressure $p$ is non-uniform owing to local variations in the cortical contractility, and due to the healing of the actin cortex after the bleb is formed. We assume that local cortical contraction is the main contribution to the change in pressure and for now prescribe the pressure of the form of a localized traveling pulse given by 
\begin{equation}
p(x,t) = \Pi \exp\left( -\frac{(x-v_{\pi} t)^2}{  x_{\pi}^2} \right),
\end{equation}
where  $\Pi$, $v_{\pi}$, and $x_{\pi}$ denote the magnitude,  traveling speed, and  width of the pulse, respectively. This form decouples the relation between actin reformation in the bleb to the changes of cortical contractility, a constraint that we will relax later on. If $v_\pi > 0$,  the pressure variations as well as the bleb boundaries $s_{l,r}$ are time dependent.  Completing the formulation of the problem for the formation and dynamics of a bleb, the boundary conditions associated with Eq.~(2) are
\begin{equation}
\frac{B}{2}  (y_{xx})^2 |_{s_l} - E_a = -\mu \left( \frac{ds_l}{dt} - v_{H-} \right)
\end{equation}
for the left boundary  and
\begin{equation}
\frac{B}{2} (y_{xx})^2 |_{s_r} - E_a = \mu \left( \frac{ds_r}{dt} + v_{H+} \right)
\end{equation}
for the right boundary where $v_{H-}$ and $v_{H+}$ are the active cortical healing speeds at the left and right boundaries of the bleb, controlled by the actomyosin interactions at either boundary. These conditions effectively assume that the dominant dissipation mechanisms are associated with the the movement of the contact lines, with $\mu$  the associated dynamic viscosity, and follow by balancing the variational derivative of the energy in Eq.~(1) with the relative velocity of the contact line due to passive mechanical forcing.  Finally, we require $\left. y\right|_{s_{l,r}} = \left. y_x\right|_{s_{l,r}} = 0$.  

Our model equations can be made dimensionless by introducing a characteristic length scale $\sqrt{B/T}$, velocity scale $T/\mu$, and energy scale $B$ so that
\begin{equation}\label{eq:beam_eq_B}
     y_{xxxx} - y_{xx} + Ky H[1-y/y_c]  = P \exp\left( -\frac{(x-v_p t)^2}{ x_p^2} \right),
\end{equation}
\begin{eqnarray}
\label{eq:boundary_eq1_B}
   \frac{1}{2} (y_{xx})^2 |_{s_l} - J &=& -\left( \frac{ds_l}{dt} - v_{h-} \right), \\
    \label{eq:boundary_eq2_B}
     \frac{1}{2} (y_{xx})^2 |_{s_r} - J &=& \left( \frac{ds_r}{dt} + v_{h+} \right).
\end{eqnarray}
with the dimensionless parameters $J,K,y_c,P,x_p,v_p,v_{h \pm}$ listed in Table~\ref{tab}. As stated earlier, in general,  $P=P(v_p,v_{h\pm})$, although we will start by assuming that $P$ is not dependent on the contractility or rate of healing.  Physiologically, the material parameters $J,K$, and $y_c$ are properties of adhesion molecules such as the Ezrin, Radixin and Moesin (ERM), proteins which are known to be key regulators of membrane-cortex interactions and signaling~\cite{ezrin}.  Thus, modifying the density of these ERM proteins in a cell will change the value of $J$, whereas modifying the sequence of the proteins may change their adhesion strength and rest length and hence $K$ and $y_c$, respectively.  On the other hand, the parameters related to the localized traveling pressure pulse, $P, x_p$, and $v_p$ are governed by the actomyosin contractile activity in the cortex; for example, modifying the motor density affects both $P$ and $x_p$ whereas modifying the actin and actin crosslinker density  changes $v_p$.  Finally, the healing speed $v_{h\pm}$ is governed by the rate of actin polymerization in the cortex. For our calculations, we assume the parameter values as listed in Table~\ref{tab}, chosen to correspond to physiological values of $B=10^{-19} \text{J}$~\cite{Charras-BioPhysJ-2008}, $T=10^{-6} \text{N/m}$~\cite{Charras-BioPhysJ-2008},$\mu=10^{-2} \text{Pa s}$, $E_a=10^{-6} \text{N/m}$, $\kappa=8 \times 10^{10} \mathrm{N/m^3}$, $l_c=5\times 10^{-9} \text{m}$, $\Pi$ ranging from 380 Pa to 475 Pa, $x_p$ ranging from $3.2 \times 10^{-7}$ m to $3.2 \times 10^{-6}$ m, and the speeds $v_p$ and $v_{h\pm}$ ranging from 0 to $2 \times 10^{-3}$ m/s.

To understand the behavior of the solutions to Eqs.~(\ref{eq:beam_eq_B})--(\ref{eq:boundary_eq2_B}), we solve them numerically with initial conditions $s_r-s_l = 2$ (or $6.3 \times 10^{-7} \text{m}$, a small patch size) and $y(x,t=0)=0$ for all $x$, using a second-order centered finite difference scheme in space and a forward Euler finite difference scheme in time.  The time-step and the grid-size used are $dt = 1 \times 10^{-4}$ and $dx = 0.025$, respectively.

\begin{largetable}
  \caption{Dimensionless parameters in the model described by Eqs.~(\ref{eq:beam_eq_B})--(\ref{eq:boundary_eq2_B}) and their values.}
  \begin{center}
  \begin{tabular}{cccc}
Symbol & Description & Dimensional description & Value\\
$J$ & membrane-cortex adhesion energy & $E_a/T$ & 1\\
$K$ & membrane-cortex spring constant & $\kappa B/T^2$ & 5000 \\
$y_c$ & critical length of membrane-cortex spring & $l_c/\sqrt{B/T}$ & 0.02\\
$P$ & magnitude of the cytoplasmic pressure & $\Pi \sqrt{B/T^3}$ & 100-150\\
$x_p$ & width of the pressure pulse & $x_{\pi}/\sqrt{B/T}$ & 1--10\\
$v_p$ & speed of the pressure pulse & $\mu v_{\pi}/T$ & 0--20\\
$v_{h\pm}$ & cortex-membrane healing speed & $\mu v_{H\pm}/T$ & 0--20\\
  \end{tabular}
  \end{center}
  \label{tab}
\end{largetable}

\begin{figure}
\includegraphics[width=\columnwidth]{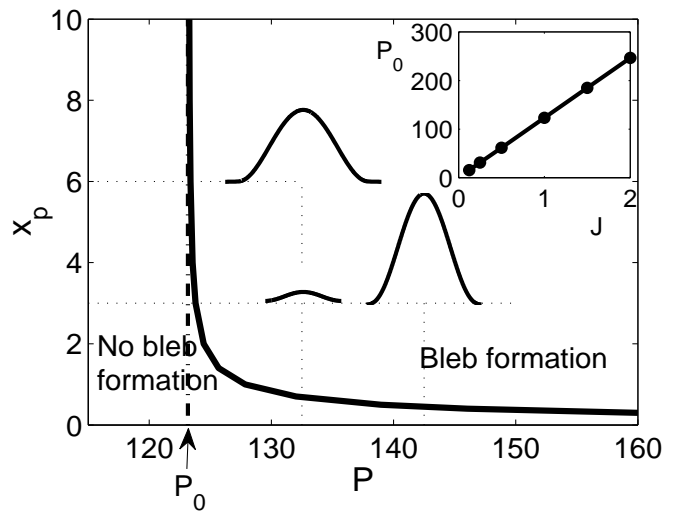}
\caption{Phase diagram of stationary bleb formation when $v_p=v_{h\pm}=0$.  Critical values of the pressure width $x_p$ and the pressure magnitude $P$ necessary for bleb formation are found to obey $x_p \sim (P - P_0)^{-\alpha}$ with $P_0=123.2$ and $\alpha=0.59$.
When only large values of $x_p > 3$ are considered (where bending energy can be neglected), the exponent approaches unity, namely $\alpha=0.73$, as previously predicted~\cite{Charras-BioPhysJ-2008}.
The inset shows numerically calculated $P_0$ as a function of adhesion energy $J$ with  $y_c$ kept constant.
These numerical results agree well with the prediction $P_0 = 2J/y_c$.  Several bleb shapes are also shown.
}
\label{fig:Fig_1}
\end{figure}

\emph{Results and Discussion --- Stationary blebs:}
We first consider the static case when $v_p = v_{h\pm} = 0$, and study the nucleation and growth of a bleb as a function of the scaled pressure magnitude $P$ and the its spatial width $x_p$. A bleb forms when membrane-cortex adhesive bonds are broken over a finite length, \emph{i.e.}, $y \geq y_c$ for some $x \in (s_l,s_r)$. When the pair of values of $x_p$ and $P$ are below a threshold, a bleb cannot form, but if a bleb can be successfully nucleated, it will grow into a symmetric stationary bleb.  The phase diagram of stationary bleb nucleation in $P- x_p$ space is shown in Fig.~\ref{fig:Fig_1}, with a phase boundary separating values of these parameters that allow for blebs from those that do not given by the curve $x_p \sim (P - P_0)^{-\alpha}$, in terms of a critical pressure $P_0$ and exponent $\alpha$, which depends on the membrane flexural rigidity $B$. We find that $\alpha \rightarrow 1$ when $x_p \gg 1$, or equivalently, $B/T \ll x_\pi^2$ in dimensional terms, \emph{i.e.}, when the membrane flexural rigidity $B$ is small, consistent with prior scaling predictions~\cite{Charras-BioPhysJ-2008} where bending energy was neglected in studying bleb nucleation. In this limit, the scaling theory \cite{Charras-BioPhysJ-2008} also predicts that the critical pressure for bleb nucleation $P_0 = 2J/y_c$, a relation we also validate numerically, as shown in the inset of Fig.~\ref{fig:Fig_1}. Sample bleb shapes shown in Fig.~\ref{fig:Fig_1} confirm that both the width and the height of a stationary bleb increase with $P$ and with $x_p$, as expected.

\begin{figure}
(a)\includegraphics[width=3.8cm]{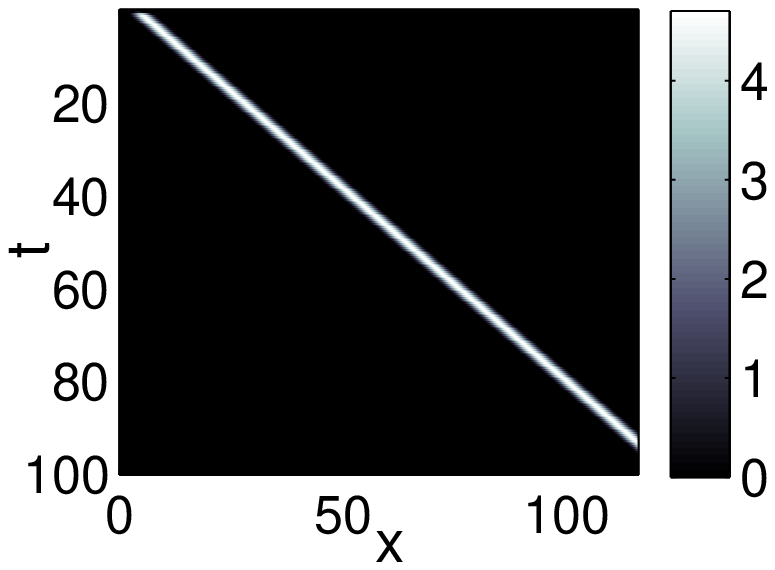}
(b)\includegraphics[width=3.8cm]{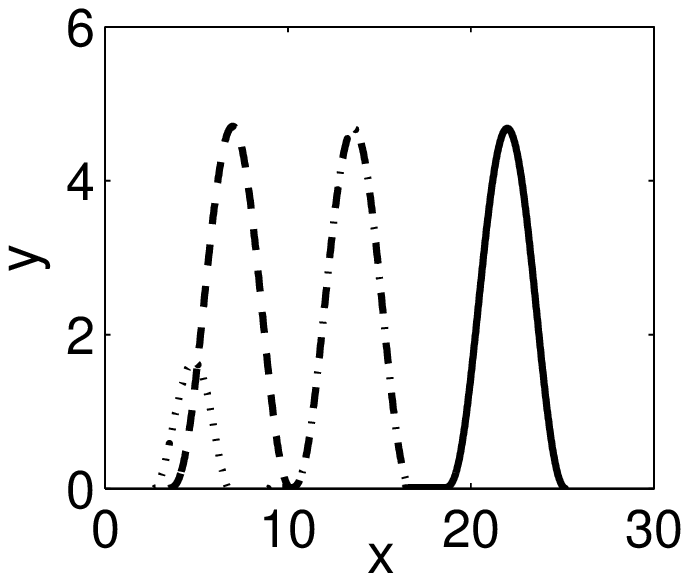}
(c)\includegraphics[width=8.2cm]{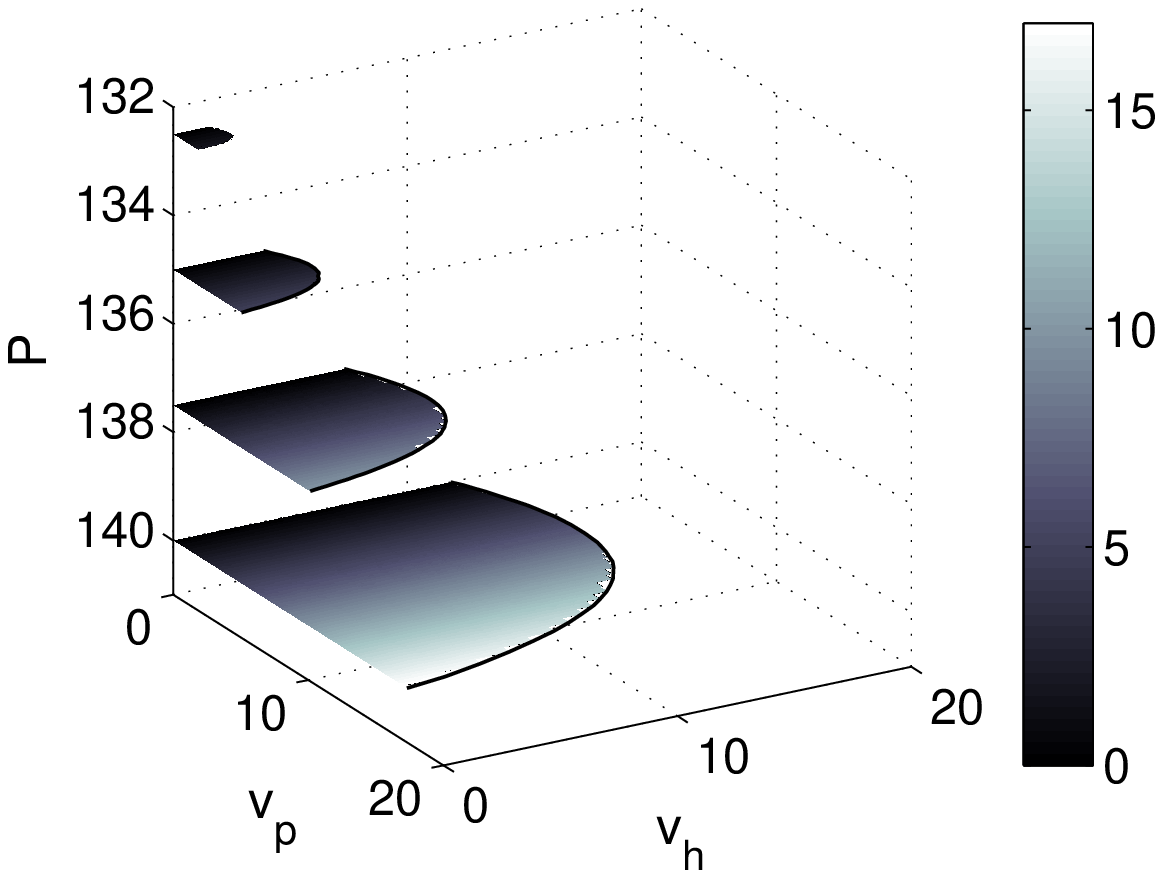}

\caption{(a) Kymograph showing a traveling bleb obtained from Eqs.~(\ref{eq:beam_eq_B})--(\ref{eq:boundary_eq2_B}) with $v_p=1.2$ and $v_h=0.6$ showing   that the traveling speed is constant; (b) Snapshots of instantaneous bleb shapes at $t= 1$ (dotted), $t=2.5$ (dashed), $t=8$ (dashed-dotted), and $t=15$ (solid).    The parameter values used were $P = 130.8$, $x_p = 9.0$, $v_p=1.2$, $v_h=0.6$ and the remaining parameter values as listed in Table~\ref{tab}. (c) Phase diagram demarcating steadily traveling and decaying phases in terms of the speed of pressure change $v_p$, speed of cortical healing $v_h$, and magnitude of pressure pulse $P$.
Traveling regimes (shaded regimes) for $P$ = 132.5, 135, 137.5, and 140 are plotted.
The traveling speed of a bleb is indicated by the colorbar.
The solid lines mark the boundaries between the two phases on the $P$ planes.
When $P$ is increased, the boundary between the steadily traveling and decaying phases is shifted outwards.}
\label{fig:Fig_5}
\end{figure}

\emph{Traveling blebs:}
We now consider the case  $v_p > 0$ and $P > P_0$ when blebs become non-stationary.  For convenience, we will set $v_{h+}=0$ and henceforth talk only of the relative healing speed $v_h = v_{h-}$.  A sample numerical calculation of  a traveling bleb is shown in Fig.~\ref{fig:Fig_5}(a-b), obtained by solving Eqs.~(\ref{eq:beam_eq_B})--(\ref{eq:boundary_eq2_B}) for parameter values that allow for bleb nucleation.   We see that a bleb first grows and moves in the direction of $v_p$; its shape eventually equilibrates and it travels  at a constant speed given by $v_p$.  However, not all traveling blebs can be sustained.  In some instances, blebs cannot be formed or retract as they travel, even if $P>P_0$.  The traveling bleb decays when either (i)  $v_h$ is large, so that the trailing edge catches up with the leading edge of the bleb, or (ii)  $v_p$ is large so that  the leading edge has a tearing rate that lags behind the location of the pressure pulse. To understand this, we consider the boundary conditions Eqs.~(\ref{eq:beam_eq_B})--(\ref{eq:boundary_eq2_B}) in a frame moving with speed $v_p$.  We see that increasing $v_p$ has the same effect as increasing the adhesion energy at the leading edge while increasing $v_h$ has the same effect as increasing the adhesion energy at the trailing edge.  Since a bleb can be sustained at equilibrium only when $P > P_0$, where $P_0$ is an increasing function of the adhesion energy $J$, a high value of either $v_p$ or $v_h$ will shift $P_0$ to a larger value, \emph{i.e.}, $P_0 \sim 2(J + v_{p,h})/y_c$, by such an extent that the bleb eventually decays. In other words,  there is a critical value of the healing speed $v_{p,h} \sim P - 2J/y_c$ above which blebs become extinct.  These results are summarized in Fig.~\ref{fig:Fig_5}(c), where we show two types of non-stationary blebs in the 3-dimensional phase space of $P$, $v_p$, and $v_h$, those that are steadily traveling and those that decay.

When a bleb travels with a steady shape and speed, our numerical simulations show that this speed is equal to $v_p$ regardless of the value of $v_h$, deviating from $v_p$ only in the initial stage of bleb growth.  Furthermore for a given traveling speed, the shape of the bleb is determined by $v_h$; a symmetric bleb exists only when $v_p=v_h/2$.  To understand this, we consider Eqs.~(\ref{eq:beam_eq_B})--(\ref{eq:boundary_eq2_B}) in a moving frame with speed $v_p$.  Then, Eq.~(\ref{eq:beam_eq_B}) becomes time-independent and the speeds of the boundaries in the moving frame are
$    ds_l/dt =  - \left(y_{xx}\right)^2/2 |_{s_l} + J - v_p + v_h  $ and
$    ds_r/dt =  \left(y_{xx}\right)^2/2 |_{s_r} - (J + v_p )$.
The steady state solution of Eq.~(\ref{eq:beam_eq_B}) with the above boundary equations is a traveling
bleb with speed $v_p$ and boundary curvatures
$    \left(y_{xx}\right)^2/2|_{s_l} = J - v_p + v_h$ and
$    \left(y_{xx}\right)^2/2|_{s_r} = J + v_p$.
Therefore, it can be seen that the bleb shape is symmetric when $v_p = v_h/2$.

\begin{figure}
\includegraphics[width=\columnwidth]{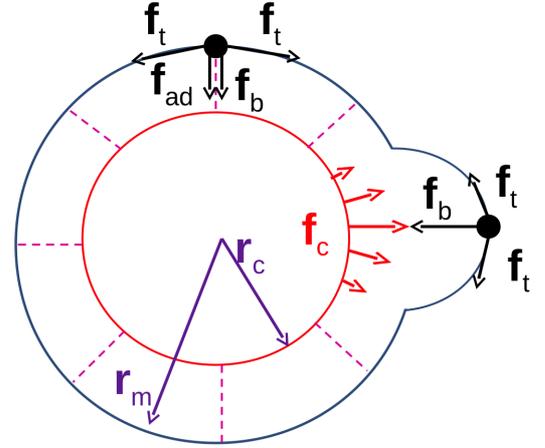}
\caption{Schematic of the model incorporating nonlocal pressure.  The position of the cell membrane ${\bf r_m} $ is solved from Eqs.~(\ref{eq:Stokes})-(\ref{eq:force}) with the forces due to membrane bending ${\bf f}_b$, membrane tension ${\bf f}_t$, membrane-cortex adhesion ${\bf f}_{ad}$ as shown.  The inner circle of radius ${\bf r_c}$ denotes the cortex contracting with ${\bf f}_c$, and the dashed lines denote membrane-cortex adhesion.
}
\label{fig:scheme}
\end{figure}

\emph{The role of a nonlocal pressure:}
In Eq.~(\ref{eq:beam_eq_B}), we assumed that the localized  perturbation  for the pressure field is independent of other variables. In reality, it is coupled to the cortical contraction and healing as well as the shape of the cell membrane, as suggested by many experiments \cite{Charras-Nature-2005}.  To account for this  coupling, we consider the interaction of the intra- and  extra-cellular fluid with the cell membrane and solve for the pressure field instead of prescribing it as an independent parameter. For relative simplicity, we assume that the other parameters associated with local cortical contraction and actin reformation, namely $v_p, x_p$, and $v_h$, remain fixed. In the context of this minimal hydrodynamic coupling model, we consider a two-dimensional cell immersed in an incompressible viscous fluid, with the same viscosity as  the cytoplasm, and enclosed by an elastic membrane uniformly adhered to a permeable but rigid actin cortex through soft Hookean springs; a schematic of the model is shown in Fig.~\ref{fig:scheme}.

Initially, the cell membrane and the cortex are assumed to be circular, with the membrane (cell) given by ${\bf r_m} = r_{m0} {\bf \hat{r}}$  and the rigid cortex given by ${\bf r_c} = r_{c0} {\bf \hat{r}}$.  The membrane-cortex adhesive springs are uniformly distributed and uniformly stretched.  This configuration gives rise to a uniform positive intracellular pressure.  Next, we assume that there is a slow global cortical contraction of the cell prescribed as $r_c(t) = r_{c0} \left[1-G\left(1-\exp(-t/\tau)\right)\right] $ where $G$ is the magnitude of global contraction and $1/\tau$ characterizes the speed of contraction. This global contraction not only generates a high uniform intracellular pressure that assists the local contractile force in nucleating and expanding a bleb, but also facilitates bleb volume growth. A bleb  nucleates when the active cortical contraction becomes large enough, modeled as a contractile body force
\begin{equation}\label{eq:contraction}
   {\bf f}_c(\theta,t) = L \exp\left[-A\left(1 - \cos\left(\theta - \frac{v_c t}{r_{c0}}\right) \right)\right] {\bf \hat{r}},
\end{equation}
of magnitude $L$ and size $\sim 1/A$, that translates with speed $v_c$ in the angular direction. 

The presence of a localized body force ${\bf f}_c$ results in a non-uniform pressure field $p(\mathbf{r})$ and velocity field ${\bf u}(\mathbf{r})$ in the fluid,
 which can be deduced from the  two-dimensional scaled Stokes equation and the equation of continuity
\begin{eqnarray}\label{eq:Stokes}
    &&\nabla p = \nabla^2 \mathbf{u} + \mathbf{f}, \\
   \label{eq:divergence_free}
    &&\nabla \cdot \mathbf{u} = 0.
\end{eqnarray}
We parameterize the cell membrane in the deformed configuration $\mathbf{r}_{m}(\zeta) = (x(\zeta),y(\zeta))^\mathrm{T}$ with $\zeta = [0,L_m]$
where $L_m$ is the perimeter of the membrane.
The motion of the membrane at position $\mathbf{r}_{m}$ is determined from the no-slip boundary condition imposed on the Stokes equation, \emph{i.e.},
\begin{equation}\label{eq:no_slip}
   \mathbf{u}_{m} = \frac{d \mathbf{r}_{m}}{dt}.
\end{equation}
The permeable rigid cortex that supports the cell membrane is assumed to be in a position such that at any time $t$, the net force experienced by the cortex is zero. Furthermore, a spring is assumed to break if the spring energy exceeds the membrane-cortex adhesion energy $J$ (i.e. if the length of the spring is greater than the critical length $y_c$); this detachment of membrane from the cortex results in the nucleation of a bleb. The total body force $\bf{f}(\mathbf{r})$ in Eq.(\ref{eq:Stokes}) is the sum of the forces from  membrane bending, membrane tension, membrane-cortex adhesion, and cytoskeletal contraction, \emph{i.e.} and is given by
\begin{eqnarray}
  {\bf f} = && \int_0^{L_m} \left[{\bf f}_b(\mathbf{r}_m) + {\bf f}_t(\mathbf{r}_m) + {\bf f}_{ad}(\mathbf{r}_m) \right] \delta(\mathbf{r}-\mathbf{r}_m)\, d\zeta \nonumber \\
  \label{eq:force}
  && + \int_0^{L_c} {\bf f}_c(\mathbf{r}_c) \delta(\mathbf{r}-\mathbf{r}_c) \, d\zeta,
\end{eqnarray}
where $L_c$ is the perimeter of the cortex and $\delta(\mathbf{r})$ is the two-dimensional Dirac delta function. In terms of the bending energy density $E_b = \gamma^2/2$ with curvature $\gamma = (x'y''-y'x'')(x^2 + y'^2)^{-3/2}$, the stretching energy density $E_t = \epsilon^2/2$ with strain $\epsilon = \sqrt{x'^2+y'^2} - 1$, and the membrane-cortex spring constant $K$, the bending force, tensile force, and adhesive force can be evaluated in a straightforward way.
We then use a boundary integral method with regularized Stokeslets \cite{Cortez-SIAM-2001} to solve the resulting equations Eqs.~(\ref{eq:Stokes})--(\ref{eq:force}), while the shape of membrane is determined using a forward Euler scheme.


This allows us to study bleb growth and dynamics by following the membrane-cortex detachment as a function of the cortical pressure, and track the bleb boundaries until membrane-cortex adhesion reformation, in accordance with Eq.~(\ref{eq:boundary_eq1_B}). We set the membrane-cortex adhesion reformation to start at one boundary after a prescribed time interval $\tau_r$ from the time when local cortical contraction occurs to prevent membrane-cortex adhesion immediately after the initial membrane-cortex detachment and  allow for bleb growth; $\tau_{r}$ plays a similar role as the initial patch size $s_r - s_l|_{t=0}$ in Eqs.~(\ref{eq:boundary_eq1_B})--(\ref{eq:boundary_eq2_B}) in nucleating the bleb.

\begin{figure}
(a)\includegraphics[width=3.8cm]{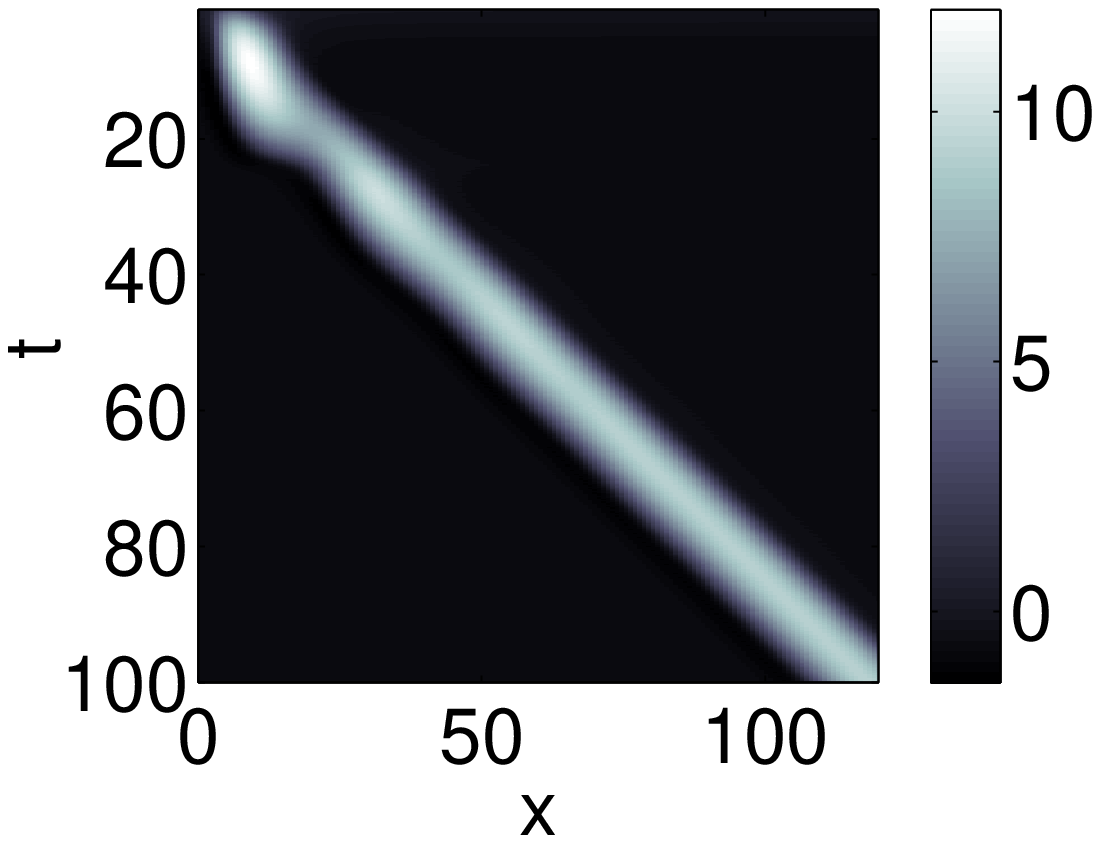}
(b)\includegraphics[width=3.8cm]{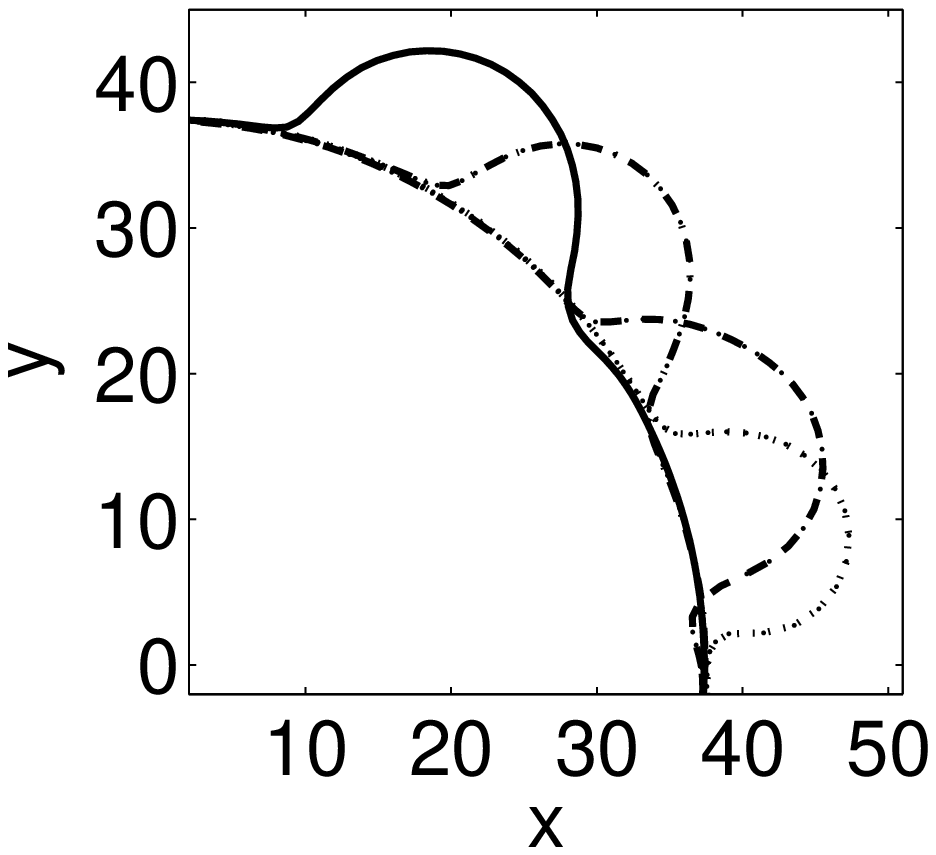}
(c)\includegraphics[width=8.2cm]{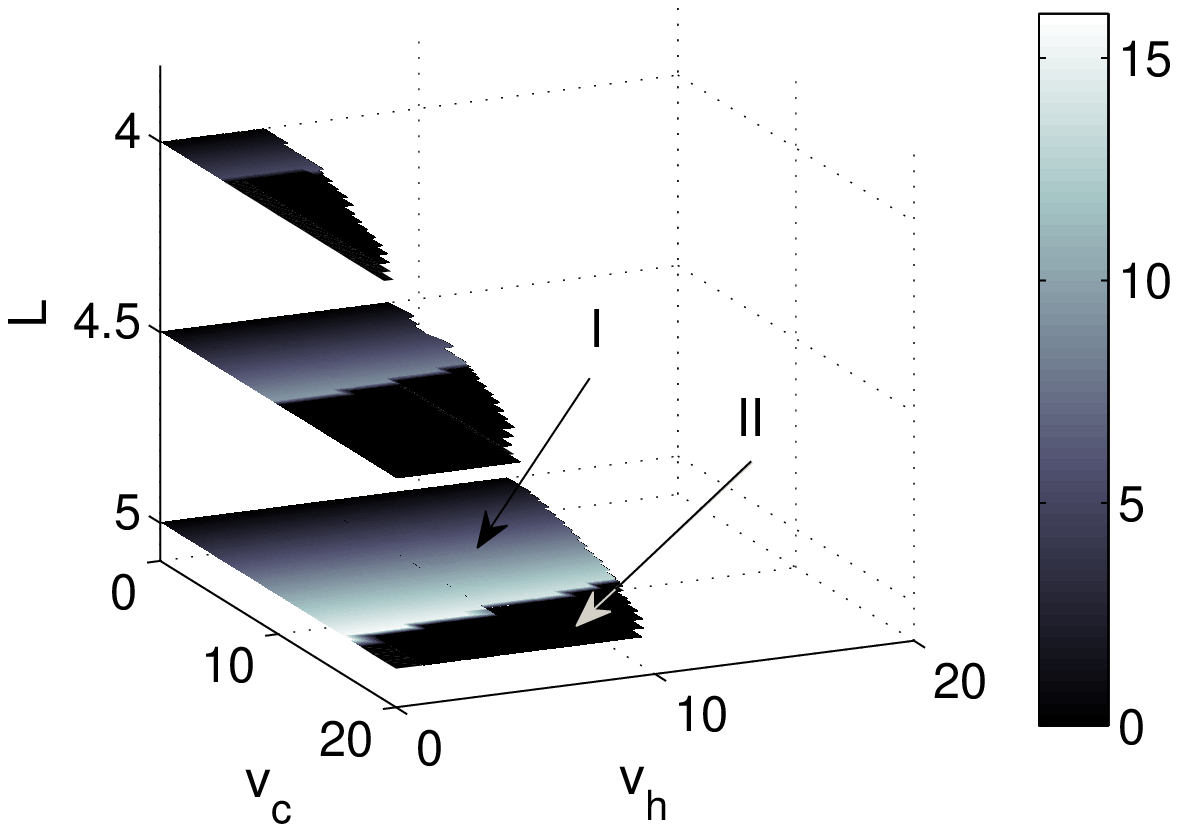}
\caption{(a) Kymograph showing a traveling bleb obtained from the Stokes model, with $v_c=1.2$ and $v_h=2.0$; (b) Snapshots of instantaneous bleb shapes at $t = 2$ (dotted), $t=12$ (dashed), $t=22$ (dashed-dotted) and $t=32$ (solid). The parameter values used were $J = 1$, $K = 20$, $L = 4$, $A = 400$, $\tau_{r}$ = 0.25, $v_c=1.2$ and $v_h=2.0$. (c) Phase diagram demarcating steadily traveling and decaying phases in terms of the speed of  localized contraction $v_c$, speed of cortical healing $v_h$, and magnitude of local contractile force $L$.
Traveling regimes (shaded regimes) for $L$ = 4, 4.5, and 5 are plotted.
A bleb is successfuly formed only when $L > 3.35$.
The traveling speed of a bleb is indicated by the colorbar.
Two traveling regimes are identified: (i) regime in which traveling speed equals to $v_c$ (regime I characterized by low $v_c$) and
(ii) regime in which traveling speed is independent of $v_c$ but is dependent on $v_h$ (regime II characterized by high $v_c$).
In the $v_h$ dependent regime, the traveling speed of a bleb is low but increases with increasing $v_h$. 
As $v_c$ further increases, this $v_h$ dependent regime gets narrower and eventually vanishes at a high enough $v_c$. 
When $L$ is increased, the boundary between the steadily traveling and decaying phases is shifted outwards.}
\label{fig:Fig_travel}
\end{figure}

Our enhanced non-local model also leads to steadily traveling blebs as shown in Fig.~\ref{fig:Fig_travel}(a-b).   Here, a bleb travels along with the local contraction speed $v_c$, equivalent to the speed of the prescribed pressure pulse motion $v_p$ in our first linear model.  In Fig.~\ref{fig:Fig_travel}(c), we show the region in phase space spanned by $v_c$ and $v_h$ for which steady traveling occurs.  We find two steady traveling regimes. In  regime I, shown in Fig.~\ref{fig:Fig_travel}(c)), the traveling speed of a bleb equals  the speed of localized contraction $v_c$ and we find good agreement in phase space with the linear model as shown in Fig.~\ref{fig:Fig_5}(c). In regime II, shown  in Fig.~\ref{fig:Fig_travel}(c)), the traveling speed depends on the healing speed $v_h$.  This occurs when the localized contractile body force is relatively small but its speed $v_c$ is relatively large. For a given rate $v_h$ at which actin polymerization and membrane healing occurs at the belt edge, the transition from regime I to regime II takes place when the contraction speed $v_c$ is so high such that the tearing rate at the leading edge fails to catch up with the local contractile force. This causes the bleb to lag behind the cortical contraction so that the bleb is then dominated by healing.  
In a finite domain with conserved volume, the combination of the nonlocal pressure and active contraction allows such a bleb to be sustained via circus movements.  However, in an infinite domain, these blebs will decay and eventually  vanish. If $v_c$ is increased even further, a bleb cannot be sustained even in a closed geometry, and there is a transition from the traveling regime II to the extinction phase, as for the linear model shown in Fig.~\ref{fig:Fig_5}(c).

\emph{Conclusion ---}
Our minimal description of the many complex phenomena associated with the onset, growth, dynamics, and extinction of blebs takes the form of a simple theory for the deformations of a thin membrane partially attached to a substrate, captured in Eqs.~(\ref{eq:beam_eq_B})--(\ref{eq:boundary_eq2_B}). We find that bleb growth and bleb motion can be modeled by detachment of the cell membrane from the underlying actin cortex through the competition between the membrane bending energy and the membrane-cortex adhesion energy, coupled with a localized pressure term. When the pressure pulse is time independent, a bleb will grow and equilibrate into a stationary bleb. However, a change in cytoplasmic pressure induced by local cortical contraction can drive the bleb to move, and lead to active blebs that travel steadily. When the pressure pulse associated with cortical contraction is an independent function, we find our theory is linear and our quantitative results are consistent with and complement earlier scaling predictions \cite{Charras-BioPhysJ-2008}. Our theory also allows us to characterize the formation and motion of blebs in terms of a phase diagram that might be useful in experimental settings.

More realistically,  the actively generated fluid pressure is coupled to membrane deformation, local cortical contraction as well  the cortical healing dynamics. To understand this,  we modified our model to account for the hydrodynamic coupling between pressure variation and membrane deformation. Our simulations show that the simple linear model is in agreement with the coupled hydrodynamic model when the localized cytoskeletal contractile body force is large.
In this case local contraction dominates  healing and the relative contribution to the pressure from hydrodynamic causes is small, so that the uncoupling of pressure from the other variables is reasonable. However, when the contractile forces are relatively small, we get active blebbing waves that persist because of a combination of passive hydrodynamic and active contractile and boundary effects, consistent qualitatively with prior experiments. Our study sets the stage for bleb driven motility, since blebs lead to asymmetric forces so that dynamics of blebs can lead to the motion of whole cells by combining the effects described here with the dynamics of adhesion of the whole cell.


\acknowledgments
We gratefully acknowledge discussions with Mike Sheetz, and partial support from the Harvard Kavli Institute for Nanobio Science and Technology, the MacArthur Foundation (LM) and the National University of Singapore through a Distinguished Visiting Professorship (LM). 

\bibliography{reference}

\end{document}